\documentclass{aastex631}

\shorttitle{T CrB: the end of the super-active phase}
\shortauthors{U. Munari}

\begin{document}

\title{The "super-active" accretion phase of T CrB has ended}

\author[0000-0001-6805-9664]{Ulisse Munari}
\affiliation{INAF Astronomical Observatory of Padova, 36012 Asiago (VI), Italy}



\begin{abstract}

The symbiotic recurrent nova T CrB erupted for the second and last recorded
time in 1946.  Following the outburst, the accretion rate onto its WD has
remained rather low with only occasional and minor flaring episodes, until in late
2014 it entered a "super-active" phase (SAP) that peaked in April 2016: the
flux radiated by Balmer lines increased by two orders of magnitude, accompanied
by the appearance of strong HeI, HeII, and many other emission lines. 
Following the sharp maximum, the intensity of the emission lines has been steadily
decreasing, reaching back the pre-SAP levels by mid-2023.  The end of SAP is
also confirmed by the drop of $B$-band brightness to pre-SAP conditions and
the simultaneous re-appearance of a large-amplitude flickering.  This suggest
that the accretion disk has emptied from the extra material that has driven
the "super active" state and has completed its transfer onto the WD, setting
the stage for a new and probably imminent nova eruption.
 
\end{abstract}



\section{Introduction} \label{sec:intro}

T CrB is a very famous recurrent nova \citep[eruptions recorded in 1866 and
1946;][and references therein]{1964gano.book.....P} and is also a
symbiotic binary by harboring a red giant (RG) as the donor star to the
massive white dwarf (WD) companion.

The life-cycle of a symbiotic binary as outlined by
\citet{2019arXiv190901389M}, is characterized by long accretion phases
interspersed by shorter periods during which the material accumulated on the
surface of the WD is burned nuclearly.  If the accreted shell is not
electron degenerate, the burning proceeds in thermal equilibrium for
decades/centuries until most of the hydrogen fuel in the shell is consumed,
the burning finally quenches down, and a new long-lasting phase of accretion
initiates the next cycle (examples are V4368 Sgr, HM Sge, and V1016 Cyg). 
When the accreted shell is instead electron degenerate, the nuclear burning
proceeds explosively resulting in a nova outburst, with most of the shell
expelled in the process and the residual nuclear burning on the WD extinguishes in a
few weeks/months, after which accretion resumes and a new cycle begins.  In
addition to T CrB, other well known symbiotic recurrent novae are RS Oph,
V3890 Sgr, and V745 Sco.

Traditionally, accretion in symbiotic stars has been treated as a smooth
process relatively stable over long periods of time
\citep[eg.][]{1986syst.book.....K}.  This approach has progressively changed
in favor of a highly-episodic interpretation of the accretion process,
characterized by brief periods of (very) high accretion rates in-between
longer intervals spent at much lower mass-transfer rates
\citep[eg.][]{2020ApJ...902L..14L,2021MNRAS.505.6121M}.
 
\citet{2016NewA...47....7M} has called attention to the fact that starting
with 2015, T CrB entered a "super-active" accretion phase (SAP),
characterized by a much brighter accretion disk as the result of a greatly
enhanced mass-flow through it and then toward the central WD.  The accretion
level attained during SAP largely exceeded any other experienced by T CrB
since the 1946 eruption.  By noting that a similar event preceded the 1946
nova outburst, \citet{2016NewA...47....7M} concluded that SAP is probably
announcing a new and imminent eruption of T CrB, a view shared by
\citet{2023arXiv230304933S}.

\begin{figure}
\includegraphics[width=17cm]{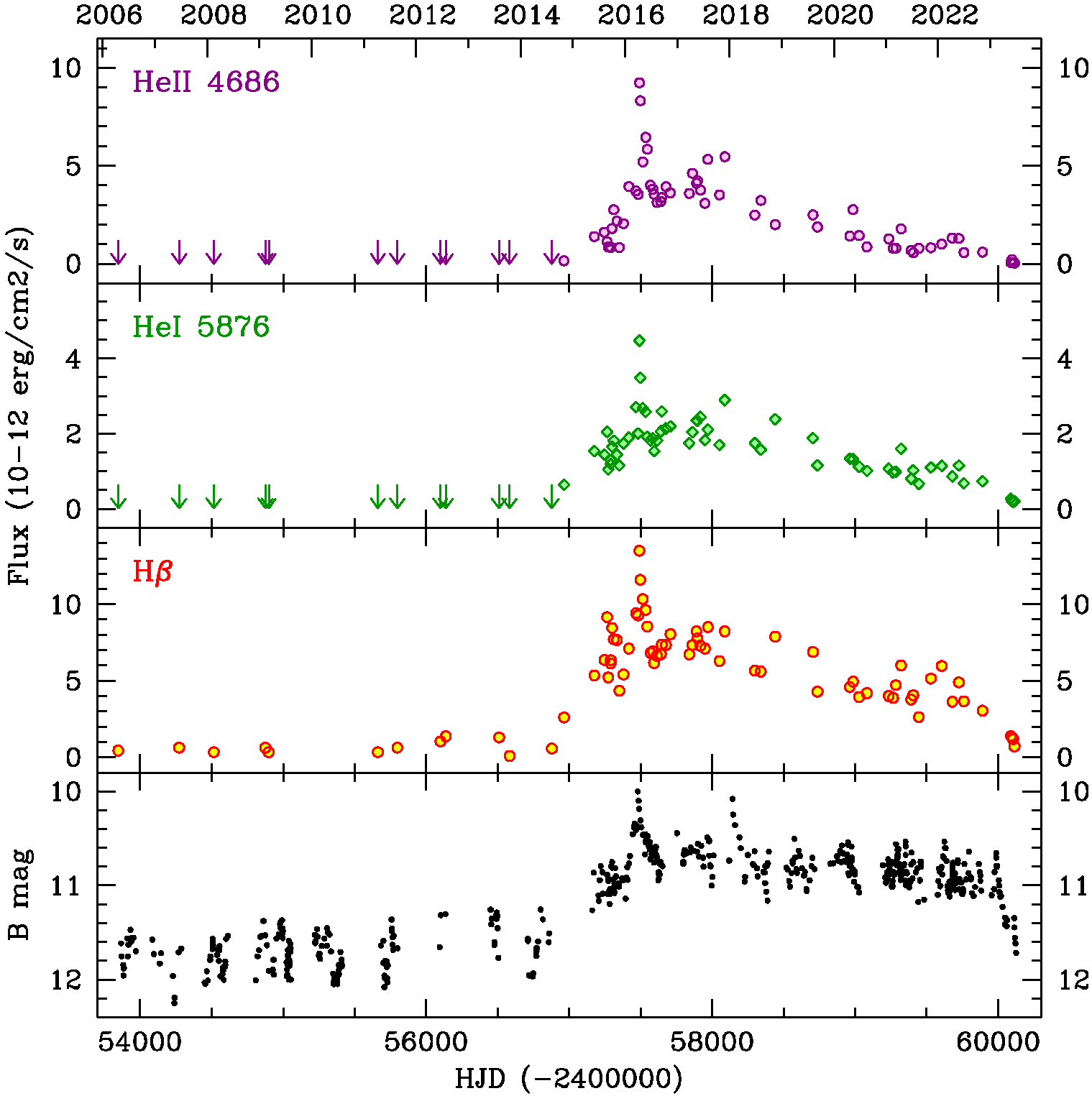}
\caption{Spectral and photometric changes prior to and during the "super-active"
accretion phase of T CrB. The bottom panel shows the $B$-band light-curve of
T CrB collected by ANS Collaboration. The panels above plot the integrated
flux of selected emission lines measured on low-resolution spectra, all obtained
with the Asiago 1.22m + B\&C (and 300 ln/mm grating). While H$\beta$ is discernible
in emission at all epochs, higher excitation/ionization lines turned on only during
the "super-active" accretion phase (the arrows point to missing-line epochs).}
\label{fig:fig1}
\end{figure}

\section{Observations} \label{sec:obs}

We have been regularly recording fluxed spectra of T CrB for the last $\sim$35 yrs,
initially with the Asiago 1.82m + B\&C and since 2006 with the
Asiago 1.22m + B\&C telescope.  For all the 1.22m spectra, we adopted a 300
ln/mm grating blazed at 5000~\AA\ that paired with a completely
UV-transparent optical train and a highly UV-sensitive CCD detector (ANDOR iDus
DU440A with a back-illuminated E2V 42-10 chip, 2048$\times$512 array, and 
13.5 $\mu$m pixel size), allows to efficiently record spectra down to the
$\sim$3100~\AA\ atmospheric cut-off imposed by the telescope 1000m altitude above sea
level.  Our 1.22m spectra of T CrB extend from 3200 to 7900 \AA\ at
2.3~\AA/pix dispersion.  In addition to being fluxed thanks to nightly
observations of spectrophotometric standard stars, their flux zero-point is
fine-tuned against (nearly-)simultaneous $B$$V$$R$ photometry, so that
the flux error anywhere in the spectra rarely exceed a few percent.  This
2006-2023 set of T CrB spectra is therefore characterized by a highly stable
instrumental set-up and robust IRAF calibration procedures, and constitutes an ideal
sample for variability studies of spectral features over long intervals of
time. A few of the spectra of T CrB here considered can be viewed in \citet{2016NewA...47....7M}.

\section{The end of the "super active" accretion phase} \label{sec:results}

To trace the evolution of T CrB along the "super active" accretion phase, we
have measured the integrated flux of a sample of emission lines on the
2006-2023 Asiago 1.22m + B\&C spectra described in the previous section. 
The selected lines are H$\beta$, HeI 5876, and HeII 4686, which are
representative of low, medium, and high excitation/ionization conditions,
respectively.  Their absolute fluxes are plotted in Figure~1 along with the
$B$-band lightcurve of T CrB as recorded by ANS Collaboration.

Prior to 2014, both HeI 5876 and HeII 4686 were not visible in emission, 
and H$\beta$ has been present but always at rather feeble levels.
During this period the $B$-band lightcurve is dominated by the ellipsoidal
distortion of the RG with superimposed the scattering due to the
large-amplitude and always present flickering \citep[eg.][and references
therein]{1998A&A...338..988Z,2010MNRAS.402.2567D}.

The start of the "super active" accretion phase in late 2014 is marked by
the sudden appearance in emission of HeI 5876 and HeII 4686, a corresponding
rise of H$\beta$ \citep[compare the spectra for
2014-11-02 and 2012-09-03 in][]{2016NewA...47....7M}, and a large increase
in $B$-band brightness caused by the rapidly brightening accretion disk. 
SAP reached its maximum in April 2016, when the flux of all emission lines
sharply peaked, as illustrated by Figure~1.  Around this epoch strong
satellite UV and thermal radio emission were also recorded
\citep{2018A&A...619A..61L,2019ApJ...884....8L}.

Following the maximum in April 2016, the flux of all emission lines has
gone steadily decreasing, at a faster pace for higher excitation/ionization
lines, and by mid-2023 they have returned to pre-SAP values, indicating that
the "super active" accretion phase is finally over.  Also the $B$-band
photometric brightness has been quickly dropping during the last few months,
while the flickering has returned to the usual large amplitude
\citep{2023ATel16023....1M} compared to the much reduced impact it had on
photometry collected around SAP maximum \citep{2016ATel.8675....1Z}.

The disappearance of emission lines, the drop in $B$-band brightness, and
the return to large amplitude flickering suggest that the accretion disk has
emptied from the extra material that driven the "super active" state and has
completed its transfer onto the WD.  The shell around the latter may
possibly still takes a little to cool and shrink down to favorable
conditions, but the stage for a new nova outburst appears now inevitably
set.

\bibliography{paper.bib}{}
\bibliographystyle{aasjournal}



\end{document}